\documentclass[sigconf]{acmart}

\AtBeginDocument{%
	\providecommand\BibTeX{{%
			\normalfont B\kern-0.5em{\scshape i\kern-0.25em b}\kern-0.8em\TeX}}}

\copyrightyear{2020}
\acmYear{2020}
\setcopyright{acmcopyright}\acmConference[CIKM '20]{Proceedings of the 29th ACM International Conference on Information and Knowledge Management}{October 19--23, 2020}{Virtual Event, Ireland}
\acmBooktitle{Proceedings of the 29th ACM International Conference on Information and Knowledge Management (CIKM '20), October 19--23, 2020, Virtual Event, Ireland}
\acmPrice{15.00}
\acmDOI{10.1145/3340531.3411993}
\acmISBN{978-1-4503-6859-9/20/10}

\usepackage{subcaption}
\usepackage{graphicx}
\usepackage{caption}

\begin{document}
\fancyhead{}

\title{E-commerce Recommendation with Weighted Expected Utility}

\author{Zhichao Xu}
\email{zhichao.xu@utah.edu}
\affiliation{%
	\institution{University of Utah}
}
\author{Yi Han}
\email{yi.han@rutgers.edu}
\affiliation{%
	\institution{Rutgers University}
}

\author{Yongfeng Zhang}
\email{yongfeng.zhang@rutgers.edu}
\affiliation{%
	\institution{Rutgers University}
}

\author{Qingyao Ai}
\email{aiqy@cs.utah.edu}
\affiliation{%
	\institution{University of Utah}
}

\begin{abstract}
	
Different from shopping at retail stores, consumers on e-commerce platforms usually cannot touch or try products before purchasing, which means that they have to make decisions when they are uncertain about the outcome (e.g., satisfaction level) of purchasing a product.
To study people's preferences with regard to choices that have uncertain outcomes, economics researchers have proposed the hypothesis of Expected Utility (EU) that models the subject value associated with an individual's choice as the statistical expectations of that individual's valuations of the outcomes of this choice.
Despite its success in studies of game theory and decision theory, the effectiveness of EU, however, is mostly unknown in e-commerce recommendation systems.
Previous research on e-commerce recommendation interprets the utility of purchase decisions either as a function of the consumed quantity of the product or as the gain of sellers/buyers in the monetary sense.
As most consumers just purchase one unit of a product at a time and most alternatives have similar prices, such modeling of purchase utility is likely to be inaccurate in practice. 
In this paper, we interpret purchase utility as the satisfaction level a consumer gets from a product and propose a recommendation framework using EU to model consumers' behavioral patterns. 
We assume that consumer estimates the expected utilities of all the alternatives and choose products with maximum expected utility for each purchase. 
To deal with the potential psychological biases of each consumer, we introduce the usage of Probability Weight Function (PWF) and design our algorithm based on Weighted Expected Utility (WEU). 
Empirical study on real-world e-commerce datasets shows that our proposed ranking-based recommendation framework achieves statistically significant improvement against both classical Collaborative Filtering/Latent Factor Models and state-of-the-art deep models in top-K recommendation.

\end{abstract}


\maketitle


\section{Introduction}

Recommendation systems are important for e-commerce as they can help consumers discover new products easily and increase the revenue of sellers and the platform.
In a typical application scenario of e-commence recommendation, products are presented to each consumer in a specific order so that the number of transactions, the profits of sellers, or, more importantly, the satisfaction of consumers could be maximized.
Different from shopping at traditional retail stores, a significant proportion of, if not all, consumer's product purchasing in e-shopping are made before the consumers can touch or try the products.
In other words, most users of e-commerce platforms make purchase decisions when they are uncertain about their potential satisfaction of the products.
Therefore, how to model this decision process and apply it for e-commence recommendation is an important research topic in both academic and industry.

In fact, how people make decisions with regard to choices that have uncertain outcomes have long been studied in the fields of economics and game theory. 
Of different theories proposed in the last few decades, the Expected Utility (EU) hypothesis~\cite{schoemaker1982expected} is considered to be the most representative and classical framework for decision theory.
The basic idea of EU is that people tend to make decisions on choices with uncertainty by first evaluating each potential outcome with their individual valuations, and then compute the subject value associated with each choice as the statistical expectations of each outcome.
Particularly in e-commerce, this means that consumer's decision on purchasing can be considered as a ranking-based process in which each consumer first estimate the payoff of each product and its alternatives based on their probabilities to produce different outcomes (e.g. satisfaction levels) according to the consumer's mental model, and then use these payoffs to generate a linear preference order among all products to find the item that gets the maximum level of satisfaction. 

Despite its intuitive theory and success in many economics and game theory applications, the effectiveness of EU in e-commerce recommendation, however, is mostly unexplored. 
While there are studies that try to combine the concept of utility with the optimization of e-commerce recommendation systems~\cite{Zhang2015, Ge2019}, most of them simply use the consumed quantity or prices/sales of each product in monetary sense to indicate the corresponding utility of each product with respect to each consumer.
As most consumers just purchase one unit of a product at a time and most products have similar prices with their alternatives, such utility assumptions often produce inaccurate consumer behavior models and, as shown in this paper, lead to suboptimal recommendation systems.

In this paper, we propose a ranking-based recommendation framework based on consumer satisfaction estimation with the Expected Utility hypothesis.
Specifically, we assume that each item purchase could produce an outcome (e.g., 5-level ratings for products on Amazon) and each outcome could lead to different level of satisfaction according to each consumer's valuation model. 
The expected utility of each item with respect to each user is the expected satisfaction levels summed over all possible outcomes.
Also, as shown by empirical studies in behavioral economics, an individual's estimation of expected utility can be affected by their psychological biases. 
For example, an irrational consumer may overestimate the probability of minor-probability events but underestimate the probability of events with large possibility to happen. 
To model such biases, we introduce the usage of Probability Weight Function (PWF) for recommendation with EU by using a non-linear function to apply weighting on the probability of each outcome for each user. 
We term the corresponding calculated utility as the Weighted Expected Utility (WEU).
The final ranked lists for each user is produced by sorting all items with their weighted expected utilities.
Experiments on real-world e-commerce dataset show that our model can achieve significant improvements over traditional and deep recommendation baselines for top-K recommendation in terms of Precision, Recall, $F_1$ measure, and NDCG.
This indicates that our framework indeed provides better consumer behavior model for e-commerce recommendation.

The contributions of this paper can be summarized as follows:
\begin{itemize}
	\item We introduce the concept of expected utility in terms of satisfaction into e-commerce recommendation. This concept, derived from real-life decision making, somehow has been overlooked by the community.
	\item We model the psychological biases of consumers by introducing probability weight functions from behavioral economics. We combine the probability weight functions with the expected utility theory and introduce the usage of weighted expected utility in terms of satisfaction.
	\item We design a recommendation framework to maximize the weighted expected utility of retrieved items.
\end{itemize}

The rest of this paper is organized as follows. 
We review the related work in Section~\ref{sec:related_work} and introduce the proposed recommendation framework in Section~\ref{sec:method}.
We show the experiment results and analysis in Section~\ref{sec:experiment} and Section~\ref{sec:results}. Finally, we summarize the paper and possible future research directions in Section~\ref{sec:conclusion}.


\section{RELATED WORK}\label{sec:related_work}
We divide the related work into five parts. First we introduce the development of classical recommendation algorithms. Then we introduce Learning to Rank (LTR), which is more relevant to the e-commerce setting where a consumer is comparing candidate items and rank them. After that we introduce the development of Economic Recommendation, which is an interdisciplinary direction of research to apply economical principles onto recommendation. In the end, we analyze the typical user behaviors in e-commerce, and introduce Prospect Theory from Behavioral Economics to study the user behaviors.

\subsection{Collaborative Filtering}
Collaborative Filtering (CF) has been well studied in the history of recommender systems. Early CF approaches consider the user-item rating matrix and conduct rating prediction with user-based \cite{Konstan1997, Resnick1994} or item-based collaborative filtering methods \cite{Sarwar2001}. User-based CF is based on the assumption that users with similar tastes for previous items would have similar preferences for new items. So the model recommends the highly ranked items by those users similar to the current user \cite{Ekstrand11, Takacs2008}. Item-based CF, or content-based CF, utilizes the features of the items to construct the user profile, and recommend new items similar to previous items the user gave good feedback. With the advancement of dimension reduction methods, Latent Factor Models are later widely adopted in recommender systems, including but not limited to singular value decomposition \cite{Sarwar2000ApplicationOD}, non-negative matrix factorization \cite{Zhang96learningfrom}, probabilistic matrix factorization \cite{Ma2008}, localized matrix factorization \cite{Zhang2013}, etc. Recent research has also extended collaborative filtering to collaborative reasoning approaches \cite{shi2020neural,chen2020neural}. In latent factor models, each user and item is learned as a latent factor representation. Well-trained latent factors are able to catch the latent features of users and items.

Many online services involve the evaluation of producers by consumers (or vice versa) through ratings. For example, on e-commerce websites such as Amazon, users are allowed to rate the purchases with a numerical star rating of 1-5; The numerical rating $r_{ij}$ reveals the satisfaction that consumer $u_i$ obtains from good $g_j$. So far, the most representative rating prediction model is the Collaborative Filtering (CF) approach based on Latent Factor Model (LFM). This approach predicts the consumer-item rating $\hat{r}_{ij}$ with the consumer/item bias and latent factors:
\begin{equation}
\hat{r}_{ij} = a+ b_{i} + l_{j} + \vec{i} \cdot \vec{j} 
\end{equation}
where $\alpha$ is the global bias, $b_i$ and $l_j$ are the consumer and item biases, $\vec{i}$ and $\vec{j}$ are the $K$-dimensional latent factors of consumer $i$ and item $j$.

\subsection{\mbox{Learning to Rank \& Top-N Recommendation}}

The state-of-the-art Learning to Rank (LTR) algorithms falls under three categories. Point-wise preference estimation works by predicting the unobserved ordinal rating values accurately \cite{li2008mcrank}. Pair-wise preference estimation works by predicting the pairwise preferences between the items \cite{bpr12, Samuels2017}. List-wise ranking estimates the preference scores for the corresponding linear order among the alternatives \cite{Suriya2016}. In our paper, we assume consumer's expected utility over a product reflects his absolute preference score. Scores of consumer-product pairs can be utilized for the construction of the top-N recommendation list. Similar to \cite{He2017}, we adopt a pairwise loss to optimize our framework.
Normalized discounted cumulative gain (NDCG) \cite{Balakrishnan2012} has been proved a good fit for Top-N recommendation tasks \cite{zhang2017joint, ai2018heterogeneous}, so we pay more attention to the NDCG performance when evaluating our framework.

\subsection{Economic Recommendation}
In e-commerce settings, a consumer makes his decisions not simply according to the average ratings of an item. Purchase behaviour is also affected by economics motives, for instance, monetary payoff, or from the psychology sense, the satisfaction level get from the purchase. For a long time, recommender system research has been focusing on the rating and ranking-related tasks, but neglected the economic motives of consumers when they purchase. Some recent research on economic recommendation has begun to take this idea into account. For example, \cite{Wang2011} first introduced utility as user's value sense in recommender systems. \cite{Zhao2015} conducted large-scale experiment for personalized promotion, e.g. customizing product price on an individual basis, to validate the consumer's sense of utility. \cite{Zhang2015} further proposed a recommendation framework to maximize the total social welfare, benefiting both the consumers and the sellers. Based on the classical microeconomics assumption that different products are inner related, \cite{Zhao2017} proposed to learn the substitutive and complementary relations between different products for the multi-product recommendation. All of the above methods interpret utility as a function of the quantity of the product. As improvement, \cite{Ge2019} focused on money efficiency, proposed to maximize the marginal utility per dollar for recommendation, to make better use of the consumers' money. \cite{ge2020learning} modeled user's personalized risk preferences for recommendation.
Although nearly all these economic theories are based on expected utility theory, none of the existing economics recommendation methods explained in detail what is expected utility in terms of satisfaction, which as we introduced previously, more suitable in e-commerce settings.

\subsection{Analysis of the User Behavior Pattern in E-commerce}
The most significant characteristic of online retail sales is the lack of physical interaction between the user and the item \cite{huseynov2016internet}. Consumer attitudes towards online shopping are usually determined by the perceived utilities \cite{hoque2015empirical} while consumers explicitly express their utilities gained by giving binary or multi-scale ratings to items \cite{ricci2011introduction}. For individual consumers, historical data of the item is critical for the purchase decision \cite{liao2012mining}. 
In this work, we assume that a user estimates the utility of the item by its historical ratings, and when making purchase decisions, he or she will rank the candidates according to the expected utilities of these items and choose among the top ones.

\subsection{Prospect Theory}
There has been extensive study in the area of behavioral economics to study the cognitive bias in decision making. The most famous theory is Prospect Theory \cite{kahneman2013prospect}. This work illustrated a series of demonstrations that people systematically violate classical Expected Utility theory when it comes to decision making, especially risk taking. Prospect theory has been used in the industry of finance \cite{benartzi1995myopic}, insurance \cite{sydnor}, and labor supply \cite{camerer1997labor}. However, after over 40 years of its original paper, there are still too few of well accepted applications. It might be tempted to conclude that although this theory is an excellent description of behaviors in experimental settings, but rather less relevant outside the laboratory. When it comes to e-commerce recommendation setting where consumers are making decisions in uncertainty, there has not been any work to apply this idea. We adopt this theory from behavioral economics, and make modifications to apply it onto e-commerce recommendation task, aiming at studying the psychological factors and cognitive biases in consumers' decision making procedure.


\section{Weighted Expected Utility for Recommendation}\label{sec:method}

In this section, we introduce our recommendation framework based on Weighted Expected Utility (WEU). We first provide some preliminary knowledge about Expected Utility hypothesis, and then describe how we estimate utility, utility distributions, and WEU for recommendation.
A summary of the notations used in this paper is provided in Table~\ref{tab:notation}.

\begin{table}[t]
	\setlength{\belowcaptionskip}{-10pt}
	\caption{A summary of notations.}
	\small
	\def\arraystretch{1.15}
	\begin{tabular}
		{| p{0.09\textwidth} | p{0.35\textwidth}|} \hline
		$i$, $j$ & An arbitrary pair of item $i$ and user $j$.   \\\hline
		$o$, $\mathcal{O}$ & An outcome ($o$) and the universal set of outcomes ($\mathcal{O}$). \\\hline
		$r$, $\hat{r}_j$ & A rating $r$ from ground truth and the reference rating point $\hat{r}_j$ for user $j$ learned by our model.\\\hline
		$u$, $p$, $w(p)$ & The utility function ($u$), the probability distribution of outcomes ($p$), and the weight of $p$ in PWF.\\\hline
		$\alpha$, $\beta$ & The parameters of $u$ in WEU.\\\hline
		$\delta$, $\gamma$, $\theta$ & The parameters of PWF.\\\hline
		$a_x$, $b_{xi}$, $l_{xj}$, $\vec{i}_x$, $\vec{j}_x$ & The parameters used to parameterize WEU parameter $x\in\{\alpha, \beta, \delta, \gamma, \theta\}$ for item $i$ and user $j$. All these parameters are learned by our model.\\\hline
	\end{tabular}\label{tab:notation}
\end{table}

\subsection{Expected Utility Hypothesis}\label{sec:EU}

Utility is an economics terminology to quantity consumer's satisfaction or fulfilment towards items. 
It is widely used to analyze the human behavior in rational choice theory \cite{Bicchieri04}.
As discussed previously, in this paper, we assume that people make decisions under uncertainty following the Expected Utility (EU) hypothesis, i.e., the subject value associated with an individual's decision is the statistical expectation of the individual's valuations of the possible outcomes.
Under the assumption of rational people, when an individual has to make a decision under uncertainty, they would make a choice with the highest expected utility \cite{coleman1992rational}.
Formally, let $u$ be the utility function for purchase decisions, then the expected utility EU for the decision of a user $j$ purchasing an item $i$ can be computed as:

\begin{equation}
EU(i,j) = \sum_{o_{ij} \in \mathcal{O}_{ij}} u_j(o_{ij})p_{i}(o_{ij})
\label{equ:EU}
\end{equation}
\vspace{-1pt}
where $o_{ij}$ is the outcome of the decision, $\mathcal{O}_{ij}$ is the universal set of potential outcomes, $u_j$ is the utility function of $j$, and $p_{i}$ is the probability distribution of outcomes for purchasing $i$.

Previous studies~\cite{Zhang2015, Ge2019} on e-commerce recommendation often formulate utility as a function of consumed quantity or prices of $i$, which limits the final utility of a purchase decision to be positive or zero.
In this paper, however, we interpret utility as the level of satisfaction that user $j$ expects to get from purchasing item $i$.
Since consumers could be either satisfied or unsatisfied with the purchase of an item, the utility of a purchase decision in our model could be positive or negative, this makes the modeling of EU more complicated as behavior economics have shown that people usually have different feelings towards decision gains and losses. 
In the rest of this section, we describe the design of our utility models and discuss how they align with well-known human decision behavior patterns from economics studies.

\begin{figure}
	\centering
	\begin{subfigure}{.25\textwidth}
		\centering
		\includegraphics[width=4.5cm]{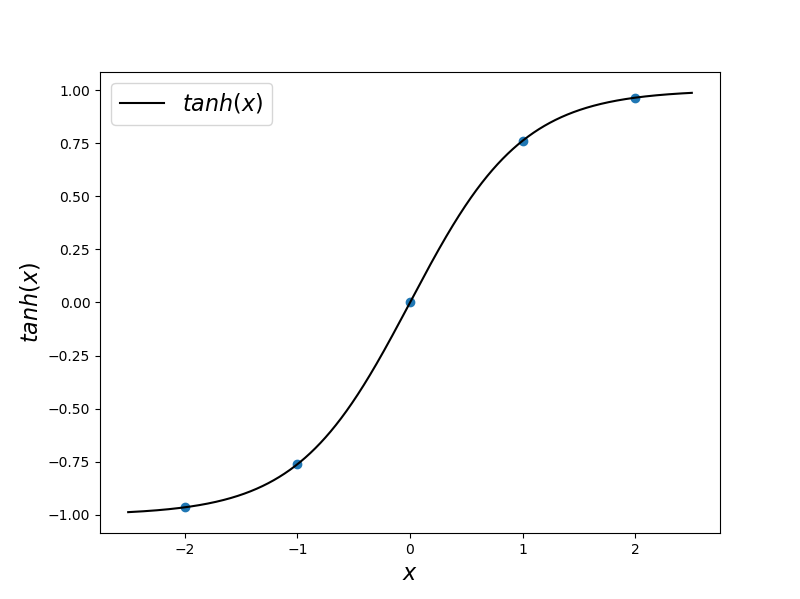} 
		\caption{$\tanh(x)$ curve}
		\label{fig:tanh}
	\end{subfigure}%
	\begin{subfigure}{.25\textwidth}
		\centering
		\includegraphics[width=4.5cm]{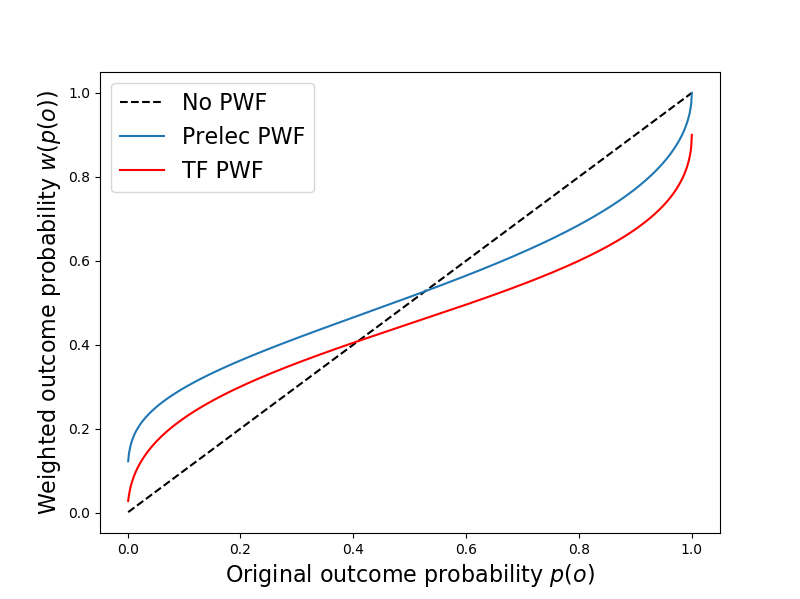}
		\caption{PWF curve}
		\label{fig:pwf}
	\end{subfigure}%
	\caption{Example curves for tanh$(x)$ and PWF. For TF, $\gamma=0.5, \delta=0.9$, and for Prelec, $\gamma=0.5, \delta=0.8$}%
	\label{fig:example}%
\end{figure}

\subsection{Model of Utility}\label{sec:utility}

We now discuss the utility function $u_j(o_{ij})$ in our model for EU estimation. 
In particular, we design $u_j(o_{ij})$ following two basic guidelines: (1) the utility of e-commerce purchases should follow the hypothesis of diminishing marginal gain, and (2) negative utility and positive utility should be treated differently.

\subsubsection{Outcome Modeling for Consumer Satisfaction}

Getting the value of positive and negative utility for each outcome is necessary for the computation of the overall expected utility of a purchase. 
Different from gambling or lottery, utility in e-commerce recommendation cannot simply be represented with a certain amount of money. 
Instead, on rating-based online shopping websites (e.g., Amazon), a more straight forward way is to represent the utility of an item for a consumer with the product ratings. 
Thus, we use the rating data on many e-commerce platforms as the indicator of consumer satisfaction to represent the value of purchase utility.

According to the decision theory~\cite{slovic1977behavioral} in behavioral economics, consumers will divide all possible outcomes as loss or gain by setting a personal reference point. 
Those outcomes smaller or worse than the reference point are classified as loss, and vice versa.
Let $r_{ij}$ be the rating of item $i$ given by user $j$, and $\hat{r}_{j}$ be the reference rating point with which user $j$ determines whether the purchase makes them satisfied or unsatisfied. 
Then we formulate the outcome of user $j$ purchasing item $i$ as
\begin{equation}
	o_{ij} = r_{ij} - \hat{r}_{j}
	\label{equ:outcome}
\end{equation}
In other words, a user would have satisfying feelings (or positive utility) when $r_{ij}$ is higher than the reference point $\hat{r}_{j}$, and unsatisfying feelings (or negative utility) otherwise. 
Note that $\hat{r}_{j}$ can be automatically learned by our model in training phase.

\subsubsection{Diminishing Marginal Utility}

In general, the marginal utility of goods or service has non-linear relationships with the consumption of that goods or service. 
Among different utility studies in economics, the law of diminishing marginal utility \cite{kauder2015history} is one of the most established theories.
The diminishing marginal utility theory states that the first unit of consumption of a good or service yields more utility than the second and subsequent units, with a continuing reduction for greater amounts. 
Mathematically, the first-order derivative of the utility function decreases as the number of consumption increases, but still greater than or equal to zero.

For the design of utility function in e-commerce recommendation, we assume that the marginal utility in terms of satisfaction also has the pattern of diminishing.
For example, the satisfaction gain from purchasing a good item comparing to a bad item should be higher than the satisfaction gain from purchasing a perfect item comparing to good items.
Therefore, the utility function should be concave in terms of gains and convex in terms of loss, namely, an S-shaped curve \cite{tversky1989rational}. 
Specifically, we adopt a popular non-linear function for utility modeling as 
$$
\tanh(x)=\frac{\exp(x)-\exp(-x)}{\exp(x)+\exp(-x)}
$$
which satisfy the law of diminishing marginal utility as shown in Figure~\ref{fig:tanh}.

\subsubsection{Differentiated Treatments for Gains and Losses}

As shown in previous studies~\cite{cohen1987experimental}, people often view gains and losses differently in decision making. 
To reflect this in the model of purchase utility, we modify the original $\tanh(x)$ function by introducing different scaling parameters on positive outcomes and negative outcomes.
Formally, the final utility function $u_j(o_{ij})$ is computed as 
\begin{equation}
u_j(o_{ij}) = \begin{cases} \alpha_j \cdot tanh(o_{ij}), & \text{if $o_{ij} \geq 0$} \\ \beta_j \cdot tanh(o_{ij}), & \text{if $o_{ij}<0$} \end{cases}
\label{equ:u}
\end{equation}
where $\alpha_j$ and $\beta_j$ are scale parameters for user $j$. 
The scaled parameters can adjust the shape of $tanh(\cdot)$, and reflect the different attitude of the consumer towards positive and negative utility.
In this paper, we learn $\alpha_j$ and $\beta_j$ automatically in model optimization, which is discussed in Section~\ref{sec:optimization}.

\subsection{Outcome Probability Distribution}\label{sec:probability}
As shown in Equation~\ref{equ:outcome}, we model consumer sanctification with respect to a product based on the rating scores provided by consumers on e-commerce platforms.  
For the sake of convenience, most e-commerce platforms only allow discrete scores for the rating of each product (e.g., 5-level rating on Amazon), which means that we need to discrete the probability distribution $p_i(o_{ij})$ on each valid rating scores for the computation of EU.
A straightforward solution for this problem is to use the empirical distribution of ratings on each item.
Given the historical rating scores $r_{i1}, \cdots, r_{iN}$ of item $i$, we estimate the probability of $o_{ij}$ using the probability of $r_{ij}$ as 
\begin{equation}
P(o_{ij}) = P(r_{ij}) = \frac{\#(r_{ij},i)}{N}
\end{equation}
where $N$ is the number of users who rated item $i$ and $\#(r_{ij},i)$ is the count of users who give the same rating to $i$ with user $j$.

\subsection{Weighted Expected Utility}\label{sec:WEU}

In practice, people often have highly personalized interpretation of the probability of possible outcomes in decision process.
In other words, different from the empirical probability distribution used in standard EU hypothesis (Eq.~\ref{equ:EU}), people usually have different feelings and behavior towards actions with small and large probabilities ~\cite{sunstein2002probability}. 
For example, most people would overrate small probability events while underrating high probability events, though the significance of such phenomenon varies depending on the personality of each person~\cite{tversky1989rational}.

To model the psychological biases in e-commerce consumer behaviors, we extended the EU hypothesis with Weighted Expected Utility (WEU) for e-commerce recommendation. 
Specifically, we adopt the probability weighting functions (PWF) to better interpret consumer's personal preferences for the probability of possible purchase outcomes.
Formally, let $w_j$ be the PWF of user $j$, then we compute the WEU of $j$ purchasing item $i$ as 
\begin{equation}
WEU(i,j) = \sum_{o_{ij} \in \mathcal{O}_{ij}} u_j(o_{ij})\cdot w_j(p_i(o_{ij}))
\label{equ:WEU}
\end{equation}
where $w_j(p_i(o_{ij}))$ be the weight of the probability of a specific outcome $o_{ij}$ when user $j$ purchases item $i$.
By introducing the personalized PWF and learning the corresponding parameters, the shape of the EU function can be adjusted based on the psychological biases of each consumer.
In this paper, we explore two potential choices for personalized PWF -- the Tversky-Fox (TF) function and the Prelec function. 
The shapes of example TF and Prelec functions are shown in Figure~\ref{fig:pwf}.

\subsubsection{Tversky-Fox Weighting}
Tversky-Fox weighting function (TF) is proposed by Tversky and Fox~\cite{TF1995} for the modeling of risk weighting under uncertainty. 
It computes the weight of probability $p$ as 
\begin{equation}
w(p) = \frac{\delta p^{\gamma}}{\delta p^{\gamma} + (1-p)^{\gamma}}
\label{equ:TF}
\end{equation}
where $\delta$ and $\gamma$ are parameters that model the relative relation between "choose" or "not choose" and the cognitive bias towards probabilities with different scales, respectively.

To extend TF for e-commerce recommendation, we personalize the parameters of TF and introduce a new parameter $\theta$ to enhance model expressive ability.
Formally, our proposed TF-based PWF for WEU computes $w_j(p_i(o_{ij}))$ as 
\begin{equation}
w_j(p_i(o_{ij})) = \frac{\delta_{j} p_i(o_{ij})^{\gamma_{j}}}{\delta_{j} p_i(o_{ij})^{\gamma_{j}} + \theta_{j}(1-p_i(o_{ij}))^{\gamma_{j}}}
\label{equ:TF+}
\end{equation}
where $0<\delta_{j}<1$, $\gamma_{j}>0$ and $0<\theta_{j}\leq1$. 
When $\theta_{j}=1$, Eq.~\ref{equ:TF} is similar to the original Tversky-Fox weighting function. 
We refer to the TF with $\theta_{j}=1$ and $0<\theta_{j}<1$ as TF and TF+, respectively.
More details about how to learn $\gamma_{j}$, $\delta_{j}$ and $\theta_{j}$ are discussed in Section~\ref{sec:optimization}.

\subsubsection{Prelec Weighting}
Prelec weighting function is proposed by Prelec~\cite{Prelec1998} as an alternative to TF, which is defined as 
\begin{equation}
w(p) = \exp \lbrace -\delta (- \ln p)^{\gamma} \rbrace
\label{equ:prelec}
\end{equation}
With the exponential function, Prelec can model better the concave or convex shape of PWF. 

Similarly, we extend the original Prelec weighting function for e-commerce recommendation as
\begin{equation}
w_j(p_i(o_{ij})) = \exp{\lbrace -\delta_{j}(-\theta_{j} \ln p_i(o_{ij}))^{\gamma_{j}} \rbrace}
\label{equ:prelec+}
\end{equation}
where $0<\delta_{j}<1$, $\gamma_{j}>0$ and $0<\theta_{j}\leq1$.
We refer the weight functions with $\theta_{j}=1$ and $0<\theta_{j}<1$ as Prelec and Prelec+. 
Again, $\gamma_{j}$, $\delta_{j}$ and $\theta_{j}$ can be learned automatically in training.

\subsection{Optimization for Top-K Recommendation}\label{sec:optimization}

In this section, we describe how we parameterize and learn the parameters in WEU for top-K recommendation.

\subsubsection{CF/LFM-based Reparameterization}\label{sec:parameterize}

In WEU, the utility functions of user $j$ purchasing item $i$ have two parameters (i.e., $\alpha$ and $\beta$) while the personalized PWF of each user $j$ have three parameters (i.e., $\delta$, $\gamma$, and $\theta$).
As demonstrated in a variety of studies~\cite{Konstan1997, bpr12, zhang2017joint, He2017}, the model of ratings or ranking scores in recommendation systems usually have three parts: the intrinsic bias of items (e.g., popularity), the intrinsic bias of users (e.g., average ratings), and the interactions between users and items.
Similarly, to model the parameters in WEU, we adopt the idea of user/item bias and latent factorizations in classical collaborative filtering and parameterize $\alpha$ and $\beta$ in Eq.~(\ref{equ:u}) as
\begin{equation}
\begin{split}
\alpha_{ij} &= a_{\alpha} + b_{\alpha i} + l_{\alpha j} + \vec{i}_{\alpha} \cdot \vec{j}_{\alpha} \\
\beta_{ij} &= a_{\beta} + b_{\beta i} + l_{\beta j} + \vec{i}_{\beta} \cdot \vec{j}_{\beta}
\end{split}
\label{equ:alpha_beta}
\end{equation}
where $a_{\alpha}$ and $a_{\beta}$ are global bias parameters, $b_{\alpha i}$ and $b_{\beta i}$ are item biases, $l_{\alpha j}$ and $l_{\beta j}$ are user bias, and $\vec{i}_{\alpha}$/$\vec{i}_{\beta}$ and $\vec{j}_{\alpha}$/$\vec{j}_{\beta}$ are the latent vector representations of item $i$ and user $j$, respectively.
Similarly, we parameterize $\delta$, $\gamma$, and $\theta$ in Eq.~(\ref{equ:TF})\&(\ref{equ:prelec}) as 
\begin{equation}
\begin{split}
\delta_{j} &= a_{\delta} + l_{\delta j} \\
\gamma_{j} &= a_{\gamma} + l_{\gamma j} \\
\theta_{j} &= a_{\theta} + l_{\theta j} \\
\end{split}
\label{equ:delta_gamma}
\end{equation}
where we ignore $b$ and $\vec{i}\cdot \vec{j}$ as the personalized PWF only depend on the user.
Note that we fix $\theta=1$ for TF and Prelec while learning $\theta$ directly in training for TF+ and Prelec+. 
In most cases, allowing $\theta$ to be learned can speed up the convergence of the final recommendation model.

\subsubsection{Discrete Choice Modeling}
In economics, discrete choice model \cite{Zhao2017} is a terminology used to describe, explain, and predict choices between two or more discrete alternatives.
Specifically, it assumes that, when making a purchase decision, a consumer always compare the product to a set of alternatives.
To apply the discrete choice model for recommendation optimization, for each item-user pair ($i$, $j$), we uniformly sampled a set of $N-1$ items from the collection in each training epoch to simulate the alternatives of $i$ for $j$ to choose. 
Together with item $i$, this forms a set of items $\Omega_{ij}$ with $N$ items for user $j$ to choose.

In this paper, we model the probability of item $i$ being chosen by user $j$ from the candidate set $\Omega_{ij}$ with the WEU of each item. 
Also, inspired by the \textit{random utility models} (RUMs) from economics \cite{Manski1977}, we inject a small noise $\epsilon_i$ to the actual value of WEU for each item to improve the robustness of our model.
Formally, let $P(i,j)$ be the probability of purchasing $i$, then we have
\begin{equation}
P(i,j) = \frac{exp(WEU(i,j)+\epsilon_i)}{\sum_{k\in\Omega_{ij}} exp(WEU(k,j)+\epsilon_k)}
\label{equ:purchase_probability}
\end{equation}
where $\epsilon_k$ is randomly sampled based on a Gaussian Distribution with mean and standard deviation as 1 and 1, respectively.

Finally, we optimize our model by maximizing the log likelihood of observed item-user purchases in the training set $\mathcal{T}$ as
\begin{equation}
\begin{split}
\mathcal{L}(\mathcal{T}) &= \sum_{(i,j)\in \mathcal{T}}\log P(i,j) - \lambda ||\Phi||^2 \\
&= \sum_{(i,j)\in \mathcal{T}}\log\frac{exp(WEU(i,j)+\epsilon_i)}{\sum_{k\in\Omega_{ij}} exp(WEU(k,j)+\epsilon_k)} - \lambda ||\Phi||^2
\end{split}
\label{equ:final_loss}
\end{equation}
where $\Phi$ is the set of all model parameters to be learned in the training process.

\section{Experimental Setups}\label{sec:experiment}
We evaluate the proposed framework based on real-world e-commerce datasets. In this section, we introduce the design and setup of the experiments.

\subsection{Description of Dataset}
To evaluate the proposed recommendation framework, we use the consumer transaction data from Amazon review datasets \cite{Ruining2016}.
Amazon review datasets have user transactions information including user ID, item ID, item rating, purchase timestamp as well as item metadata such as price, related items, etc. 
In this paper, we conduct experiments on three categories of the Amazon review datasets, i.e., \textit{Baby}, \textit{Movies}, and \textit{Electronics}. The original dataset is huge and sparse, similar to previous work \cite{kang2018self}, we filter out the users and items with fewer than 10 interactions. To better compare with previous economic recommendation method \cite{Ge2019, Zhang2015}, we only use items with prices.
The basic statistics of each category are shown in Table~\ref{tab:statistics}.

\begin{table}
	\centering
	\setlength{\abovecaptionskip}{3pt}
	\caption{Basic Statistics of Datasets}
	\resizebox{0.475\textwidth}{!}{
	\begin{tabular}{ | c | c | c | c | c |}
		\hline
		Dataset & \#Users & \#Items & \#Interactions & \#Sparsity \\
		\hline
		Movies & 25,431 & 10,470 & 726,857 & 0.273\% \\
		\hline
		Electronics & 40,983 & 16,286 & 556,227 & 0.083\% \\
		\hline
		Baby & 23,894 & 19,834 & 166,459 & 0.035\% \\
		\hline
	\end{tabular}
	}
	\label{tab:statistics}
\end{table}

In order to create the training, validation, and test sets for experiments, for each user, we sort their transaction records according to the corresponding purchase timestamps. 
Then, for each category, we split the transaction records in chronological order into training, validation, testing sets by 3:1:1. 
More specifically, for each user, the first 60\% items he purchased and are used for training, next 20\% are used for validation, and last 20\% for testing.

\subsection{Baselines}
We compare our proposed models with state-of-the-art recommendation algorithms, including both traditional methods and deep learning methods. 
To show the effectiveness of the weighted expected utility framework, we also include baselines that incorporate economic user models for recommendation algorithms as well as a simplified version of our framework without utility weighting.

\subsubsection{Classical Recommendation Algorithms}
In this paper, we include two classical recommendation algorithms in our experiments -- the Collaborative Filtering (CF) model~\cite{Ekstrand11} based on Latent Matrix Factorization (LFM) and the Bayesian Personalized Ranking (BPR) model~\cite{bpr12}. 

The Collaborative Filtering model based on Latent Matrix Factorization (CF-LFM) is a classic recommendation model that builds user and item representations with matrix factorization.
In training, we used the rating matrix of training data to learn CF-LFM, and, in testing, we rank items according to the predicted rating of each user for each candidate based on CF-LFM.

The Bayesian Personalized Ranking (BPR) model extend standard collaborative filtering methods for top-K recommendation by computing a pairwise loss from implicit feedback (purchased or not). 
It directly produces a ranking score for each user-item pair so that we could recommend items by sorting them with their scores.

\subsubsection{Deep Learning based Algorithm}
In this paper, we use NCF, short for Neural Collaborative Filtering \cite{He2017}, a state-of-the-art recommendation algorithm based on the structure of deep neural network (DNN) \cite{He2017, zhang2017joint, wang2014collaborative} as our baseline. 
Compared to traditional recommendation methods, DNN-based models are more sufficient in capturing complex user-item interactions. 
In our experiment, we use the NCF that fuses the Generalized Matrix Factorization (GMF) and MLP. 
Same as in CF-LFM, in training, the rating matrix of the training data is used to train NCF.
In testing, we recommend items based on their predicted ratings for each user.

\subsubsection{Economic Recommendation Algorithms}
We compare our proposed framework with two other economic recommendation algorithms -- the Total Surplus Maximization (TSM) model \cite{Zhang2015} and the Maximizing Marginal Utility per Dollar (MUD) model \cite{Ge2019}.

The Total Surplus Maximization \cite{Zhang2015} model is one of the first recommendation algorithms constructed based on economic principles and utility theory. 
It considers prices and purchase quantity information in the modeling of purchase utility.

The Maximizing Marginal Utility per Dollar \cite{Ge2019} is an economic recommendation algorithm that focused on marginal utility provided by the item. 
Similar to TSM, it also utilizes the price of the product as a factor for decision making.

\subsubsection{EU, a simplified version of our proposed method}:
In order to show the effectiveness of utility probability weighting, we create a simplified version of our model with only expected utility theories. 
In other words, the simplified model ranks the products' utilities without the weight function. 
Specifically, we use the following equation instead of WEU in Eq.~\ref{equ:WEU} to rank items:
\begin{equation}
EU(i,j) = \sum_{o_{ij} \in \mathcal{O}_{ij}} u_j(o_{ij}) \cdot p_i(o_{ij})
\label{equ:EU_baseline}
\end{equation}
We refer to this simplified version of our model as EU.

\subsection{Experimental Settings}

\subsubsection{Evaluation Protocol}
To evaluate our model's robustness under sparse settings, we use the evaluation protocols similar as in \cite{zheng2019deep} and \cite{zhang2017joint}. Specifically, for each user, we randomly sample 1,000 negative items, and rank them with the ground truth items of the user. We use standard top-K recommendation metrics including Precision, Recall, $F_1$, and NDCG.

\subsubsection{Hyperparameters Setting}
In our experiment, we tuned four hyperparameters: latent vector size, the weight of regularizers, the SGD optimizer's momentum, and learning rates. 
All the baselines are implemented using 64 as latent vector size. We implement our models using latent vector size in $[16, 32, 64, 128]$. For our models and all the baselines, the weight of regularizers is searched from 1e-4 to 1; the learning rate is searched from 1e-4 to 5e-1; and $[0, 0.1, 0.5]$ is used for the choice of momentum. 
We also discuss the impact of choosing different latent vector size in Section~\ref{sec:results}.
Our datasets and implementation will be available at \footnote{https://github.com/zhichaoxu-shufe/E-commerce-Rec-with-WEU}

\begin{table*}[h]
\centering
\setlength{\abovecaptionskip}{3pt}
\caption{Summary of the recommendation performance. We evaluate for ranking (\textbf{P}, \textbf{R}, \textbf{$F_1$}, \textbf{NDCG}), $K$ is the length of the recommendation list. Weighted Expected Utility models' improvement against best baselines are significant at p = 0.001}
\resizebox{\textwidth}{!}{
\begin{tabular}{|ccccccccccccc|}
\hline
\multicolumn{1}{|c|}{Dataset}  & \multicolumn{12}{c|}{Baby}  \\ 
\hline
\multicolumn{1}{|c|}{Measures} & \multicolumn{3}{c|}{Precision(\%)} & \multicolumn{3}{c|}{Recall(\%)} & \multicolumn{3}{c|}{$F_1$ Measure(\%)} & \multicolumn{3}{c|}{NDCG} \\ 
\hline
\multicolumn{1}{|c|}{K} & 1 & 5 & \multicolumn{1}{c|}{10} & 1 & 5 & \multicolumn{1}{c|}{10} & 1 & 5 & \multicolumn{1}{c|}{10} & 1 & 5 & 10 \\ 
\hline
CF & 3.6872 & 2.6580 & 1.9297 & 2.6134 & 9.2453 & 13.6695 & 2.5535 & 2.6355 & 3.1929 & 0.0368 & 0.0283 & 0.0599 \\
BPR & 4.1855 & 3.0413 & 2.4834 & 2.9667 & 10.7771 & 17.5992 & 3.4722 & 4.7439 & 4.3526 & 0.0419 & 0.0472 & 0.0836 \\
NCF & 4.2424 & 3.0976 & 2.3936 & 3.0105 & 10.9384 & 17.9409 & 3.5218 & 4.8280 & 4.2237 & 0.0424 & 0.0712 & 0.0921 \\
EU & 3.5191 & 2.0537 & 1.3864 & 2.4612 & 7.1175 & 9.5493 & 2.8966 & 3.1876 & 2.4213 & 0.0352 & 0.0583 & 0.0667 \\
TSM & 4.2320 & 3.1318 & 2.3582 & 2.9522 & 10.8354 & 17.9571 & 3.4781 & 4.8591 & 4.1689 & 0.0423 & 0.0655 & 0.0873 \\
MUD & 4.4155 & 3.2327 & 2.4533 & 3.1542 & 11.0385 & 18.0045 & 3.6797 & 5.0009 & 4.3182 & 0.0441 & 0.0825 & 0.1074 \\

\hline
Prelec & 5.4239 & 3.4427 & 2.7367 & 3.9267 & 12.2423 & 19.4050 & 4.5554 & 5.3741 & 4.7969 & 0.0542 & 0.0979 & 0.1225 \\
Prelec+ & 5.5077 & \textbf{3.4611} & 2.7601 & 3.9950 & \textbf{12.3775} & 19.5469 & 4.6309 & \textbf{5.4095} & 4.8372 & 0.0551 & 0.0990 & 0.1237 \\
TF & 5.3319 & 3.3699 & 2.6233 & 3.9337 & 12.0667 & 18.7022 & 4.5273 & 5.2685 & 4.6012 & 0.0533 & 0.0965 & 0.1194\\
TF+ & \textbf{5.6123} & 3.4360 & \textbf{2.7681} & \textbf{4.0548} & 12.2592 & \textbf{19.6077} & \textbf{4.7081} & 5.3676 & \textbf{4.8513} & \textbf{0.0561} & \textbf{0.0992} & \textbf{0.1234} \\
\hline
\end{tabular}
}
\vspace{1pt}

\resizebox{\textwidth}{!}{
\begin{tabular}{|ccccccccccccc|}
\hline
\multicolumn{1}{|c|}{Dataset}  & \multicolumn{12}{c|}{Electronics}  \\ 
\hline
\multicolumn{1}{|c|}{Measures} & \multicolumn{3}{c|}{Precision(\%)} & \multicolumn{3}{c|}{Recall(\%)} & \multicolumn{3}{c|}{$F_1$ Measure(\%)} & \multicolumn{3}{c|}{NDCG} \\ 
\hline
\multicolumn{1}{|c|}{K} & 1 & 5 & \multicolumn{1}{c|}{10} & 1 & 5 & \multicolumn{1}{c|}{10} & 1 & 5 & \multicolumn{1}{c|}{10} & 1 & 5 & 10 \\ 
\hline
CF & 4.8733 & 3.1126 & 2.3305 & 2.1601 & 6.8984 & 10.3312 & 2.9934 & 4.2895 & 3.8034 & 0.0487 & 0.0634 & 0.0755 \\
BPR & 6.9134 & 3.8314 & 2.7275 & 3.0652 & 9.4927 & 12.0904 & 4.2473 & 5.2801 & 4.4503 & 0.0691 & 0.0821 & 0.0949 \\
NCF & 7.0766 & 3.8912 & 2.8548 & 3.3408 & 9.7398 & 12.9790 & 4.5388 & 5.5608 & 4.6802 & 0.0708 & 0.0854 & 0.1004 \\
EU & 7.7422 & 3.7518 & 2.6992 & 4.0481 & 9.2089 & 12.7918 & 5.3164 & 5.3315 & 4.4578 & 0.0742 & 0.0832 & 0.0952 \\
TSM & 6.8372 & 3.8876 & 2.7191 & 3.1304 & 9.3106 & 12.4737 & 4.2945 & 5.4849 & 4.4649 & 0.0684 & 0.0793 & 0.0921 \\
MUD & 7.9327 & 3.8961 & 2.9346 & 4.0789 & 9.6705 & 13.0584 & 5.3876 & 5.5544 & 4.7922 & 0.0793 & 0.0875 & 0.1034 \\

\hline
Prelec & 8.2058 & 4.1476 & 2.8770 & \textbf{4.2381} & 10.2306 & 14.0423 & \textbf{5.5894} & 5.9023 & 4.7745 & 0.0821 & 0.0943 & 0.1031\\
Prelec+ & 8.1594 & 4.3633 & 3.1794 & 4.1650 & 10.8030 & 15.5740 & 5.5149 & 6.2160 & 5.2807 & 0.0815 & \textbf{0.0937} & 0.1096 \\
TF & 8.2058 & 4.3544 & 3.1713 & 4.1897 & 10.8118 & 15.4727 & 5.5471 & 6.2084 & 5.2637 & 0.0821 & 0.0937 & 0.1104 \\
TF+ & \textbf{8.2814} & \textbf{4.3721} & \textbf{3.2225} & 4.2108 & \textbf{10.8420} & \textbf{15.7745} & 5.5829 & \textbf{6.2314} & \textbf{5.3517} & \textbf{0.0828} & 0.0945 & \textbf{0.1105} \\
\hline
\end{tabular}
}

\vspace{1pt}

\resizebox{\textwidth}{!}{
\begin{tabular}{|ccccccccccccc|}
\hline
\multicolumn{1}{|c|}{Dataset}  & \multicolumn{12}{c|}{Movies}  \\ 
\hline
\multicolumn{1}{|c|}{Measures} & \multicolumn{3}{c|}{Precision(\%)} & \multicolumn{3}{c|}{Recall(\%)} & \multicolumn{3}{c|}{$F_1$ Measure(\%)} & \multicolumn{3}{c|}{NDCG} \\ 
\hline
\multicolumn{1}{|c|}{K} & 1 & 5 & \multicolumn{1}{c|}{10} & 1 & 5 & \multicolumn{1}{c|}{10} & 1 & 5 & \multicolumn{1}{c|}{10} & 1 & 5 & 10 \\ 
\hline
CF & 2.9372 & 2.4754 & 2.1963 & 0.5466 & 2.3021 & 4.0858 & 0.9211 & 2.3852 & 2.8567 & 0.0239 & 0.0315 & 0.0375 \\
BPR & 4.7782 & 3.8743 & 3.4246 & 0.8896 & 3.6032 & 6.3687 & 1.4992 & 3.7344 & 4.4536 & 0.0478 & 0.0504 & 0.0601 \\
NCF & 4.8279 & 4.1065 & 3.7268 & 1.3126 & 4.1935 & 7.9233 & 2.0640 & 4.1495 & 5.0692 & 0.0483 & 0.0557 & 0.0644\\
EU & 4.5456 & 2.5417 & 1.6920 & 1.1147 & 2.9058 & 3.3653 & 1.7904 & 2.7116 & 2.2518 & 0.0454 & 0.0361 & 0.0363 \\
TSM & 4.7028 & 3.9748 & 3.4622 & 1.0753 & 3.4624 & 6.2623 & 1.7504 & 3.7009 & 4.4591 & 0.0470 & 0.0530 & 0.0597 \\
MUD & 5.1040 & 3.5641 & 2.9043 & 1.3190 & 4.5166 & 7.8493 & 2.0963 & 1.9254 & 4.2398 & 0.0510 & 0.0484 & 0.0552 \\

\hline
Prelec & 6.2758 & 4.7768 & 4.2004 & 1.6170 & 5.9406 & 10.1449 & 2.5714 & 5.2957 & 5.9410 & 0.0628 & 0.0634 & 0.0766 \\
Prelec+ & 6.3971 & 4.8004 & 4.2192 & 1.6059 & 5.9417 & 10.1825 & 2.5661 & 5.3104 & 5.9662 & 0.0640 & 0.0639 & 0.0771 \\
TF & 6.2050 & 4.7737 & 4.2141 & 1.6164 & 5.9288 & 10.1175 & 2.5647 & 5.2889 & 5.9500 & 0.0620 & 0.0632 & 0.0765 \\
TF+ & \textbf{6.4233} & \textbf{4.8254} & \textbf{4.2329} & \textbf{1.6371} & \textbf{6.0032} & \textbf{10.3562} & \textbf{2.6092} & \textbf{5.3502} & \textbf{6.0095} & \textbf{0.0642} & \textbf{0.0650} & \textbf{0.0785} \\

\hline
\end{tabular}
}
\label{tab:results}
\end{table*}

\section{Results and Analysis}\label{sec:results}

In this section, we introduce our experiment results and analysis. The results of recommendation performance are displayed in Table~\ref{tab:results}. 
We refer to our models with different probability weighting function in Section~\ref{sec:WEU} as TF, TF+, Prelec, and Prelec+.
For all the baselines and our models, a latent vector size of 64 is used for the reported results.

\subsection{Recommendation Performance}
For top-k recommendation performance, we first compare the performance between traditional CF-LFM model with NCF, and then compare our models with both former economic recommendation algorithms and non-economic recommendation algorithms. 

\subsubsection{Traditional CF-LFM vs Neural CF Recommendation}
The NCF baseline is a state-of-the-art recommendation algorithm. With the complex connectivity and the nonlinearity in the neural networks, NCF is capable of properly estimating the complicated interactions between the user and the item in the latent space. 
As we can see from Table~\ref{tab:results}, on all three datasets, NCF achieves significantly better performance in terms of NDCG when compared to traditional CF (i.e., CF-LFM), BPR and EU. In Baby dataset, it outperforms EU, BPR, CF by 38.1\%, 10.2\%, 53.6\% respectively in terms of NDCG@10. 
Note that EU is a simplified version of our weighted expected utility recommendation framework and can be seen as adding a simple modification to the classical CF-LFM algorithm. 
These results confirm the arguments from previous studies that deep neural network has great potentials catching the complex user-item interactions~\cite{He2017}.

\subsubsection{Comparison with Non-economic Recommendation Algorithms}
We compare the performance of our weighted expected utility framework (WEU) with three non-economic recommendation algorithms: CF, BPR and NCF. Non-economic recommendation algorithms purely focus on catching the complex non-linear user-item interactions, while our models aim to catch the personalized user preferences and behavior models with the weighted expected utility theory.
As can be seen from Table~\ref{tab:results}, the Weighted Expected Utility model we proposed achieves better performance in terms of all four measures when compared to best non-economic recommendation baseline (i.e., NCF). Specifically, in Baby dataset, TF+ model achieves 15.64\%, 9.29\%, 14.86\%, 33.98\% improvement in P@10, R@10, $F_1$@10, NDCG@10 respectively. This indicates the value of the weighted expected utility theory in e-commerce and the effectiveness of our proposed framework in the task of ranking and constructing recommendation item lists to users. 

\subsubsection{Comparison with other economic recommendation algorithms}
In our experiments, we also compare the performance of our framework with other recommendation frameworks that incorporate economic user models, namely TSM and MUD. 
On all the datasets we tested, WEU significantly outperforms TSM and MUD by a large margin. 
More specifically, in Baby dataset, TF+ model achieves 12.83\%, 8.90\%, 12.34\%, 14.90\% improvement in P@10, R@10, $F_1$@10, NDCG@10 respectively. This suggests that the weighted utility theory used in our proposed recommendation framework can better catch and model users' purchase behaviors. 

\subsection{Analysis and Discussion}

In this section, we focus on the interpretation of the user behavior models learned in our framework and try to answer the question: \textit{does WEU actually learn useful knowledge as we expected?}
Specifically, we analyze the utility functions and probability weighting functions learned by our models on the Baby dataset from two perspectives: personalized risk attitudes and psychological biases to the probabilities.
We also provide some hyper-parameter analysis to show the effect of different parameter settings.

\begin{figure*}[]
    \setlength{\abovecaptionskip}{3pt}
    \setlength{\belowcaptionskip}{3pt}
	\centering
	\begin{subfigure}{.33\textwidth}
		\centering
		\includegraphics[width=6cm]{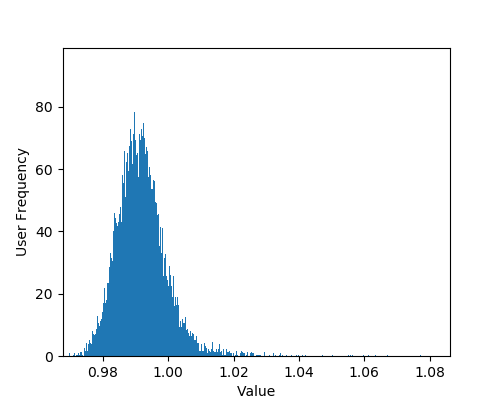} 
		\caption{$\overline{\alpha}_j$}
	\end{subfigure}%
	\begin{subfigure}{.33\textwidth}
		\centering
		\includegraphics[width=6cm]{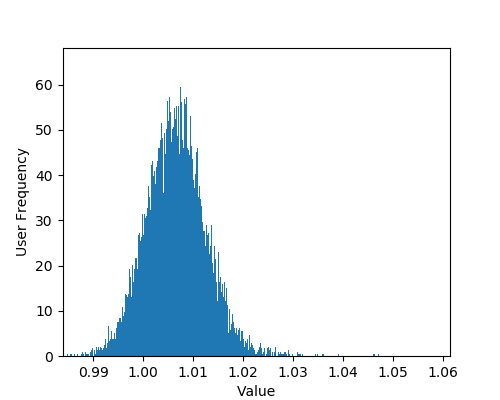}
		\caption{$\overline{\beta}_j$}
	\end{subfigure}%
	\begin{subfigure}{.33\textwidth}
		\centering
		\includegraphics[width=6cm]{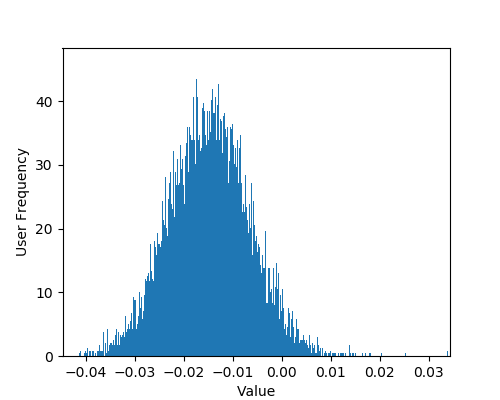}
		\caption{$\overline{\alpha}_j - \overline{\beta}_j$}
	\end{subfigure}%
	\vspace{-5pt}
	\caption{Distribution of $\overline{\alpha}_j$, $\overline{\beta}_j$ and $\overline{\alpha}_j - \overline{\beta}_j$ of all test users in Baby Dataset in Prelec+}
	\vspace{-15pt}
	\label{fig:alpha_beta}
\end{figure*}

\subsubsection{Personalized Risk Attitudes}

Classical Prospect Theory argues that most decision makers are risk-averse, which means they are more tempted to be conservative in risk taking. 
In order to verify whether such argument holds in the proposed model, we extract $\alpha_{ij}$ and $\beta_{ij}$ in Eq.~\ref{equ:u} from our Prelec+ model on the Baby dataset.
For simplicity, we average the value of $\alpha_{ij}$ and $\beta_{ij}$ for each user $j$ on all test items to indicate user's overall preferences over expected gains and losses.
Formally, we have
\begin{equation}
	\overline{\alpha}_j = \frac{\sum_{i \in \mathcal{I}}\alpha_{ij}}{|\mathcal{I}|},~~~	\overline{\beta}_j = \frac{\sum_{i \in \mathcal{I}}\beta_{ij}}{|\mathcal{I}|}
\end{equation}
where $\mathcal{I}$ is the universal set of items in the testing data.


As shown in Figure~\ref{fig:alpha_beta}, in the Baby dataset, the distribution of $\overline{\alpha}_j$ and $\overline{\beta}_j$ extracted from our Prelec+ model mostly follow the shape of normal distributions.


The distribution of $\overline{\alpha}_j-\overline{\beta}_j$ has a mean value that is negative and only a tiny proportion of users have $\overline{\alpha}_j$ larger than $\overline{\beta}_j$, which means that the utility functions of most users are left-skewed.
Also, in Figure~\ref{fig:baby_tanh}, we show Prelec+ and TF+'s utility functions based on the averaged $\overline{\alpha}_j$ and $\overline{\beta}_j$ for all users in the Baby dataset. 
In comparison with the initial setting of $tanh(x)$, the learned utility function now weight positive utility less and weigh negative utility more.
These observations meet the economic theory that most decision makers are risk-averse, and they weigh losses more than gains. 


In our experiment, we do notice that the difference between $\overline{\alpha}_j$ and $\overline{\beta}_j$ in terms of absolute value is not large.  
One possible reason is that the prospect theory is designed based on lottery, of which utility can be computed directly with money.
When using ratings to estimate user's utility in purchasing items on e-commerce websites, however, we don't have such straightforward measurements and user's sense of gain and loss in terms of purchase satisfaction could be vague.
In fact, as shown in Table~\ref{tab:results}, using the expected utility function only for recommendation (i.e., EU in Table~\ref{tab:results}) doesn't produce superior performance, which indicates that the utility function in Eq.~(\ref{equ:u}) itself is not enough for the modeling of user's mental and behavior models in recommendation.  



\subsubsection{Psychological Biases to Outcome Probabilities}

\begin{figure}
	\centering
	\begin{subfigure}{.24\textwidth}
		\centering
		\includegraphics[width=4.6cm]{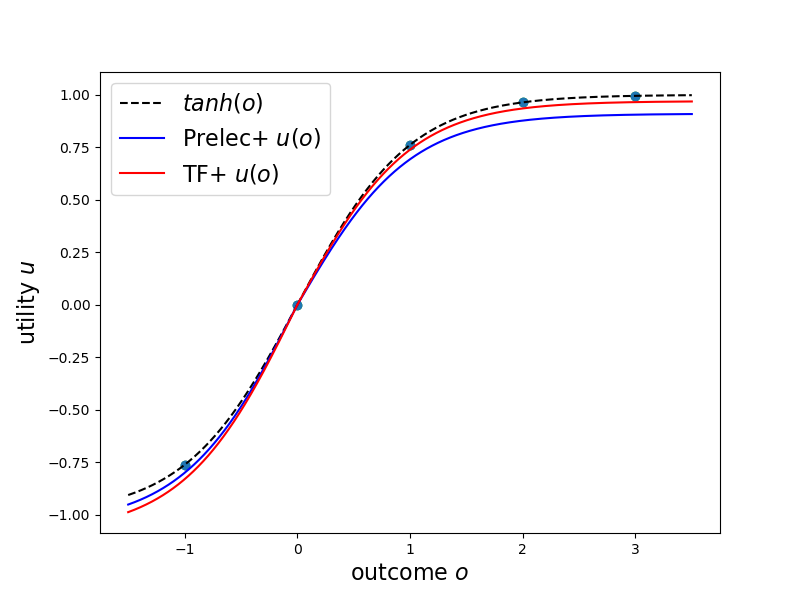} 
		\caption{Learned utility function}
		\label{fig:baby_tanh}
	\end{subfigure}%
	\begin{subfigure}{.25\textwidth}
		\centering
		\includegraphics[width=4.6cm]{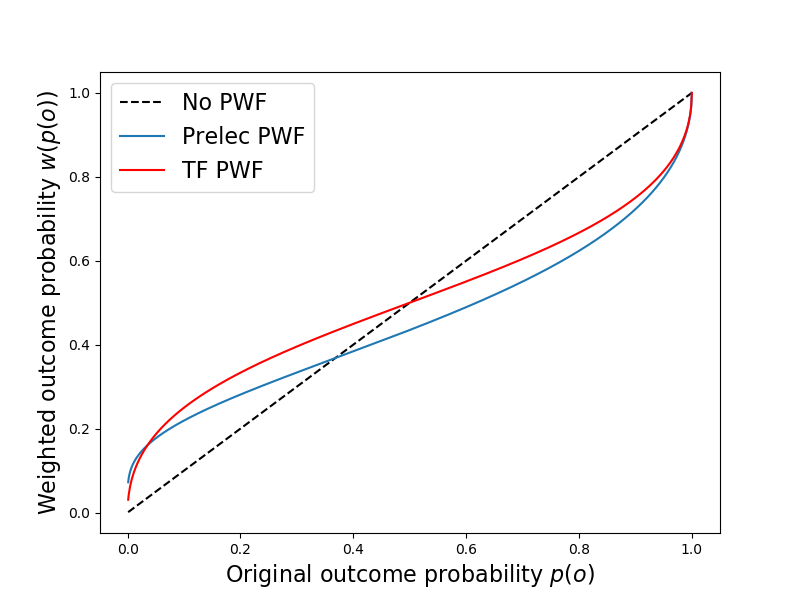}
		\caption{Learned PWF curve}
		\label{fig:baby_pwf}
	\end{subfigure}%
	\vspace{-10pt}
	\caption{Learned utility functions and PWF functions of our model in Baby Dataset.}
	\vspace{-15pt}
	\label{fig:baby_example}%
\end{figure}

Psychological study and Prospect Theory shows that risk-averse decision makers tend to overweight small probabilities to guard against losses. 
In Figure~\ref{fig:baby_pwf}, we plot the average PWF curves for Prelec+ and TF+ using the mean value of $\gamma_j$, $\delta_j$, $\theta_j$ in Eq.~(\ref{equ:TF+})\&(\ref{equ:prelec+}) for all test users in Baby dataset. 
We also show the outcome probability with no weighting for comparison.
As shown in the figure, on average, users tend to overweight small probabilities while underweight large probability events.

As shown in Table~\ref{tab:results}, on all three datasets, our proposed weighted EU framework achieves significant improvement over EU baseline. In Electronics dataset, TF+ model achieves 19.39\%, 23.32\%, 20.05\%, 16.07\% improvement in P@10, R@10, $F_1$@10, NDCG@10 respectively. 
This indicates modeling consumers' psychological bias to probabilities by introducing probability weight function can indeed help us achieve better recommendation performance.
In our experiments, we observed that extreme ratings, for example, 1 out of 5, even only a few of them, can affect the decisions of users greatly. So it is both necessary and important to study the probability biases.

As discussed in Section~\ref{sec:WEU}, we extend Prelec PWF and TF PWF by adding new parameter $\theta_j$ to adjust the weighting between "choose" and "not choose". 
In our experiments, PWF+ models consistently achieves comparable or better performance than PWF models. For example, In Baby dataset, TF+ model achieves 5.52\%, 4.84\%, 5.44\%, 3.35\% improvement in P@10, R@10, $F_1$@10, NDCG@10 against TF respectively;  Prelec+ model achieves 0.96\%, 1.10\%, 0.84\%, 0.98\% improvement in P@10, R@10, $F_1$@10, NDCG@10 against Prelec respectively. Also, adding new parameters leads to quicker convergence.
This indicates that giving more flexibility to the learning of PWF could be beneficial for the overall performance of recommender systems.

\begin{figure*}[]
    \setlength{\abovecaptionskip}{3pt}
    \setlength{\belowcaptionskip}{3pt}
	\centering
	\begin{subfigure}{.33\textwidth}
		\centering
		\includegraphics[width=6cm]{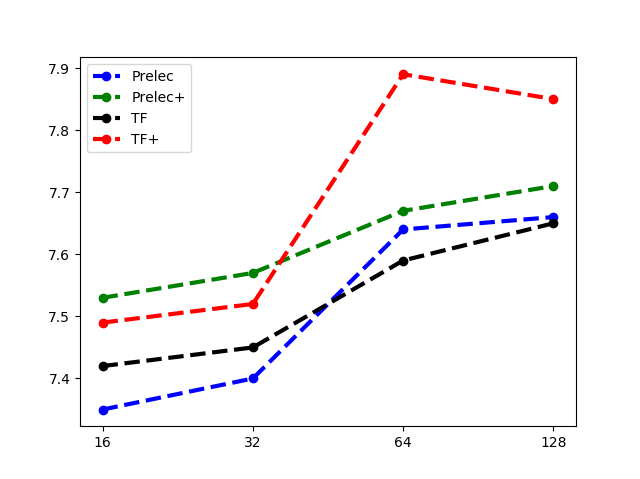} 
		\caption{Movies Dataset}
	\end{subfigure}%
	\begin{subfigure}{.33\textwidth}
		\centering
		\includegraphics[width=6cm]{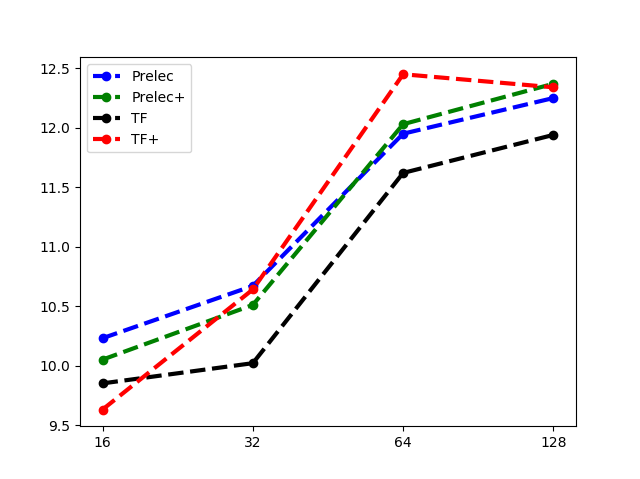}
		\caption{Baby Dataset}
	\end{subfigure}%
	\begin{subfigure}{.33\textwidth}
		\centering
		\includegraphics[width=6cm]{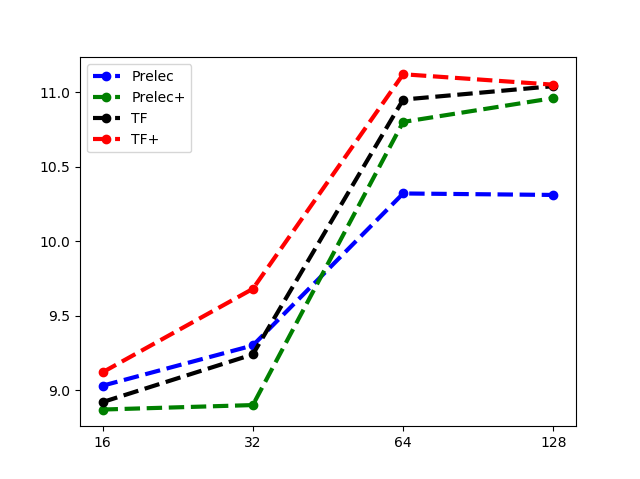}
		\caption{Electronics Dataset}
	\end{subfigure}%
	\caption{The comparison between using different latent vector sizes, Y axis is NDCG@10$\times 10^2$}
	\vspace{-10pt}
	\label{fig:latent_size}
\end{figure*}

\subsubsection{Impact of applying different latent vector sizes}
In Figure~\ref{fig:latent_size}, we show the impact of latent vector size to our models with respect to NDCG@10. 
Similar to previous studies~\cite{bpr12, Ge2019, He2017}, we observe that the performance of our models increases as latent vector size increases. 
However, after certain points, the performance improvements with larger latent sizes vanish or become smaller. 
We notice that TF+'s best latent vector size is 64, while all other three models' performance still improves when changing from 64 to 128. 
Overall, our models are generally robust with respect to the settings of latent vector sizes.


\section{CONCLUSION AND FUTURE WORK}\label{sec:conclusion}

In this paper, we propose to apply the Expected Utility theory from economics to the construction of e-commerce recommendation systems. 
Specifically, we construct a personalized utility function based on user's ratings for items and propose a weighted expected utility framework to model the decision process and psychological bias of e-commerce consumers.
Our experiments show that most consumers tend to be risk-averse as they give higher weights to negative utilities and lower weights to positive utilities.
Also, according to our analysis on the probability weighting functions, most consumers in our experiments tend to overweight outcomes with low probabilities while underweight outcomes with high probabilities.
Empirical study on real-world e-commerce dataset shows that our personalized ranking-based recommendation model achieves better performance in terms of $F_1$ Measure, NDCG when compared with non-economic algorithms, including classical CF-LFM models and state-of-the-art DNN-based models.

In this paper, we model user's purchase utility and satisfaction purely based on their ratings for the items.
Such paradigm is suboptimal as user's opinions on each purchase are reflected and affected by not just ratings, but also reviews, purchase prices, discounts, and much more.
How to jointly utilize those different categories of information to construct better user bebavior models, however, is still unclear.
In the future, we plan to explore more economic theories or user studies for the design of better e-commerce recommendation systems, and to explain the recommendations from economic perspectives \cite{zhang2014explicit,zhang2020explainable}.

\section{Acknowledgements}
This work was supported in part by the School of Computing, University of Utah. Any opinions, findings and conclusions or recommendations expressed in this material are those of the authors and do not necessarily reflect those of the sponsor.

\bibliographystyle{ACM-Reference-Format}
\balance
\bibliography{sample-base}

\end{document}